\documentstyle[twocolumn,aps,epsfig,here]{revtex}

\begin{document}

\title{Ratios of charged antiparticles to particles near mid-rapidity in Au+Au collisions \\
at $\sqrt{s_{_{NN}}} =$ 130 GeV }
\author { B.B.Back$^1$, M.D.Baker$^2$, 
D.S.Barton$^2$, R.R.Betts$^{1,6}$, R.Bindel$^7$,  
A.Budzanowski$^3$, W.Busza$^4$, A.Carroll$^2$,
M.P.Decowski$^4$, 
E.Garcia$^7$, N.George$^1$, K.Gulbrandsen$^4$, 
S.Gushue$^2$, C.Halliwell$^6$, 
G.A.Heintzelman$^2$, C.Henderson$^4$, R.Ho\l y\'{n}ski$^3$, D.Hofman$^6$,
B.Holzman$^6$, 
E.Johnson$^8$, J.Kane$^4$, J.Katzy$^{4,6}$, N. Khan$^8$, W.Kucewicz$^6$, P.Kulinich$^4$,
W.T.Lin$^5$, S.Manly$^{8}$,  D.McLeod$^6$, J.Micha\l owski$^3$,
A.Mignerey$^7$, J.M\"ulmenst\"adt$^4$, R.Nouicer$^6$, 
A.Olszewski$^{2,3}$, R.Pak$^2$, I.C.Park$^8$, 
H.Pernegger$^4$, C.Reed$^4$, L.P.Remsberg$^2$, 
M.Reuter$^6$, C.Roland$^4$, G.Roland$^4$, L.Rosenberg$^4$, 
P.Sarin$^4$, P.Sawicki$^3$, 
W.Skulski$^8$, 
S.G.Steadman$^4$, 
G.S.F.Stephans$^4$, P.Steinberg$^2$, M.Stodulski$^3$, A.Sukhanov$^2$, 
J.-L.Tang$^5$, R.Teng$^8$, A.Trzupek$^3$, 
C.Vale$^4$, G.J.van Nieuwenhuizen$^4$, 
R.Verdier$^4$, B.Wadsworth$^4$, F.L.H.Wolfs$^8$, B.Wosiek$^3$, 
K.Wo\'{z}niak$^3$, 
A.H.Wuosmaa$^1$, B.Wys\l ouch$^4$\\
(PHOBOS Collaboration) \\
$^1$ Physics Division, Argonne National Laboratory, Argonne, IL 60439-4843\\
$^2$ Chemistry and C-A Departments, Brookhaven National Laboratory, Upton, NY 11973-5000\\
$^3$ Institute of Nuclear Physics, Krak\'{o}w, Poland\\
$^4$ Laboratory for Nuclear Science, Massachusetts Institute of Technology, Cambridge, MA 02139-4307\\
$^5$ Department of Physics, National Central University, Chung-Li, Taiwan\\
$^6$ Department of Physics, University of Illinois at Chicago, Chicago, IL 60607-7059\\
$^7$ Department of Chemistry, University of Maryland, College Park, MD 20742\\
$^8$ Department of Physics and Astronomy, University of Rochester, Rochester, NY 14627\\
}
\maketitle

\begin{abstract}\noindent
We have measured the ratios of 
antiparticles to particles for charged pions, kaons and protons
near mid-rapidity in central Au+Au collisions 
at $\sqrt{s_{_{NN}}} =$ 130 GeV. 
For protons, we observe $\langle \overline{p} \rangle / \langle p \rangle
 = 0.60 \pm 0.04 (stat.) \pm 0.06 (syst.)$ in the 
transverse momentum range $0.15 < p_T < 1.0$~GeV/c. This leads to
an estimate of the baryo-chemical potential $\mu_B$ of 45~MeV,
a factor of 5-6 smaller than in central Pb+Pb collisions at 
$\sqrt{s_{_{NN}}} = 17.2$~GeV.
\end{abstract}


In this paper multiplicity ratios of antiparticles to particles for primary charged pions, kaons 
and protons are presented for collisions of gold nuclei at an energy of $\sqrt{s_{_{NN}}} = 130$~GeV. 
The data were taken with the PHOBOS detector 
during the first run of the Relativistic Heavy-Ion Collider (RHIC)
at Brookhaven National Laboratory. 
The experiments at RHIC aim at understanding the behavior of 
strongly interacting matter at high temperature and density.
Quantum chromodynamics,
the fundamental theory of strong interactions, predicts that under these
conditions a new state of matter will be formed, the
quark-gluon plasma \cite{qgp}. 

One of the most intriguing results from
heavy-ion collisions at lower energies was the observation that 
particle ratios for particles with production cross-sections 
varying by several orders of magnitude could be described in a
statistical picture of particle production assuming chemical 
equilibrium \cite{heinz,becattini,pbm}. 
This is particularly remarkable for the production rates of rare baryons 
and antibaryons containing multiple strange quarks \cite{koch}, which 
were found to be difficult to reproduce in microscopic 
hadronic transport models \cite{urqmd,rqmd}. One of the key ingredients 
for the statistical particle production picture is the 
baryo-chemical potential $\mu_B$. The 
particle ratios presented here, in particular 
$\langle \overline{p} \rangle / \langle p \rangle$,  provide 
information about $\mu_B$ at RHIC energies.

On a microscopic level, the 
$\langle \overline{p} \rangle / \langle p \rangle$ ratio
reflects the interplay between the transport of baryon number 
carried by protons and neutrons in the colliding nuclei in  momentum 
space, the production of quark-antiquark pairs and the 
annihilation of antiprotons in the final stages of the collision.
The connection between baryon number transport and the energy 
loss of the incoming nucleons \cite{busza} is under intense theoretical discussion \cite{kharzeev,vance}. 

The data reported here were collected using the PHOBOS 
two-arm magnetic spectrometer.  Details of the setup 
can be found elsewhere\cite{phobos1,phobos2,pak}. 
One arm (SPECP) of the 
spectrometer was only partially equipped with 6 layers of silicon 
sensors, providing tracking only in the field-free region close to 
the beampipe. The other arm (SPECN) 
had a total of 16 layers of sensors, providing tracking both outside and 
inside the 2~T field of the PHOBOS magnet. Particles within the geometrical 
acceptance region used in this analysis traverse 14 or 15 of the layers.
A two layer silicon detector (VTX) covering $|\eta| < 1.5$ and 25\% of 
the azimuthal angle provided additional information on the position
of the primary collision vertex.
In total, 94362 sensitive detector elements were read out, 
of which less than 2\% were non-functional. 

The primary event trigger was provided by two sets of 16 scintillator 
paddle counters covering pseudorapidities 
$3 < |\eta |< 4.5$.
Additional information for event selection was obtained from two 
zero-degree calorimeters measuring spectator neutrons. 
Details of the event selection and centrality determination 
can be found in \cite{phobosprl,judith_qm2001}. 
Monte Carlo (MC) simulations of the apparatus
were based on the HIJING event generator \cite{hijing} 
and the GEANT~3.21 simulation package, folding in
the signal response for scintillator counters and silicon sensors.

For this analysis the 12\% events with the 
highest signal in the paddle counters were selected, corresponding to 
the collisions with the highest number of participating 
nucleons $N_{part}$.
The average number of participants for the selected events 
was estimated as $\langle N_{part} \rangle = 312 \pm 10 \mbox{(syst.)}$, 
using a Glauber calculation relating $N_{part}$ to the fractional 
cross section observed in 
bins of the paddle signal \cite{phobosprl,judith_qm2001}.

As the geometrical layout of the PHOBOS detector leads to an asymmetry 
in the acceptance and detection efficiency for positively and 
negatively charged particles for a given magnet polarity, 
data was taken using both magnet polarities (BPLUS and BMINUS). 
The reproducibility of the absolute field strength was found to be 
better than 1\%, based on Hall probe measurements for each polarity
and the comparison of mass distributions for identified particles
for the two polarities.

The stability of the trigger event selection was checked using 
the number of reconstructed straight line particle tracks 
outside the magnetic field. 
The straight track multiplicities for the BPLUS and BMINUS data sets
agreed to within 0.2\%.  

\begin{figure}[t]
\centerline{
\epsfig{file=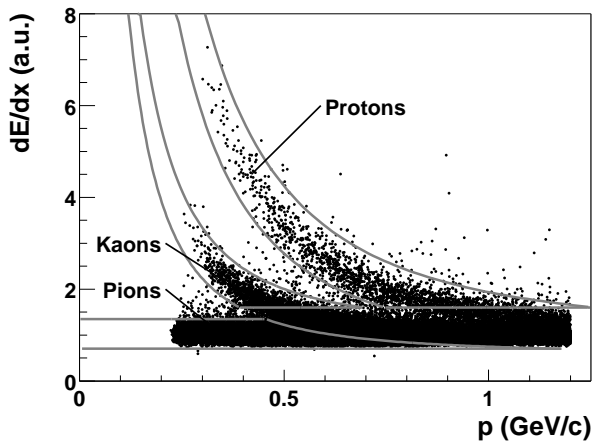,width=8.5cm}
}
\caption{Distribution of average energy loss as a function of 
reconstructed particle momentum. Three clear bands can be seen,
corresponding to pions, kaons and protons. The solid lines indicate
the cut regions for counting identified particles. }
\label{dedx_p}
\end{figure}
To optimize the precision of the vertex  and track
finding  only events with a reconstructed primary vertex position 
between -16~cm $< z_{vtx} < $ 10~cm along the beam axis were selected.
We also restricted the analysis to tracks in the central region of the 
spectrometer planes along the 45$^\circ$ axis of the experiment.
A precise knowledge of the primary vertex for 
each event is essential for this analysis, as the distance of 
closest approach of each reconstructed track with respect to the 
primary vertex ($dca_{vtx}$) is the only tool for rejecting 
background particles from decays and secondary interactions.
By requiring a consistent vertex position from the SPEC and 
VTX subdetectors, in combination with the known position of the 
beam orbit, a vertex resolution better than 
0.3~mm (RMS) in $y$ and $z$ directions and better than 0.5~mm in
$x$ direction was achieved. 

Particle tracks in the spectrometer 
were found as follows:  
tracks in the first 6 layers outside the magnetic field 
were reconstructed by a road-following algorithm. 
\begin{figure}[t]
\centerline{
\epsfig{file=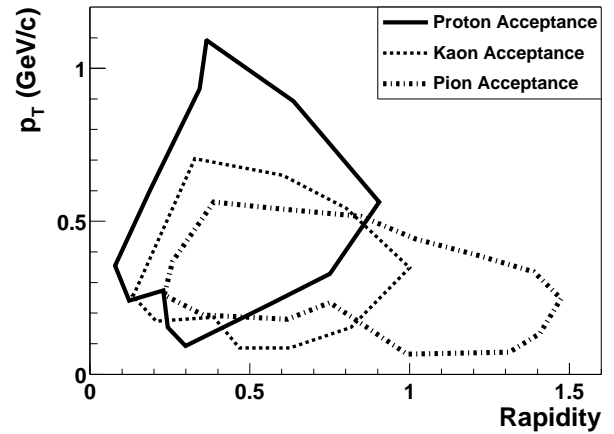,width=8.5cm}}
\caption{Acceptance of the spectrometer as a function of transverse momentum 
and rapidity for pions, kaons and protons. The acceptance is averaged over 
the accepted vertex range and azimuthal angle $\phi$.}
\label{accept}
\end{figure}
\noindent Inside 
the field, two-hit combinations on consecutive layers were 
mapped into ($1/p,\Theta$) space, where $\Theta$ is the polar
angle at the primary interaction vertex and $p$ the total momentum.
A clustering algorithm then selected combinations of matching 
hits in ($1/p,\Theta$) space, yielding track pieces inside the field.
For track pieces inside and outside the field,
the specific ionization $dE/dx$ was calculated using the 
truncated mean of the angle-corrected hit energies, discarding
the two highest energy values. 
Straight and curved track pieces 
were then matched based on $\Theta$, a fit to the combined track 
in the $yz$-plane and requiring consistency in the independently 
determined $dE/dx$ values. 

For the combined tracks the $dE/dx$ was again calculated, discarding the 4 highest 
energy values. 
Particle identification was performed using $p$ and  
$dE/dx$, which depends only on particle 
velocity. The identification cuts are shown in Fig.~\ref{dedx_p} 
with three bands corresponding to pions, kaons and protons. 
The corresponding acceptance regions for identified particles in 
transverse momentum $p_T$ and 
rapidity are shown in Fig.~\ref{accept}.
Table~\ref{table1} gives the resulting event and identified particle statistics for 
all combinations of magnet polarity, particle charge and particle mass, as 
well as the uncorrected average transverse momentum $\langle p_T \rangle$.
Using two magnetic field settings allows the determination of two statistically
independent values for each of the three particle ratios. As can
be calculated from the numbers in Table~\ref{table1}, the two values 
agree within statistical uncertainty for each of the ratios.

Preliminary PHOBOS data show a total multiplicity  of produced charged particles
$N_{ch} \approx  3500$ \cite{alan_qm2001}  for the event selection used here.
This is large compared to the average initial charge asymmetry given by 
the number of participating protons of $\langle N_{part}^p \rangle \approx 125$. The 
charge asymmetry $1 - \langle N^- \rangle /\langle N^+ \rangle$ in the 
final state will therefore be small,
so that a random contamination of the identified particle 
samples would move the observed ratios towards unity.
The contamination of the proton and antiproton samples was checked by testing 
the stability of the antiproton/proton ratio against variation 
of the particle-ID cuts shown in Fig.~\ref{dedx_p}.
Within statistical error the ratios were stable against further 
changes of the cuts.
 
For comparison with data from other experiments and theoretical calculations
several corrections to the observed particle ratios  have to be applied. 
These corrections account for particles produced
in secondary interactions, loss of particles due to absorption in detector material,
and feeddown particles from weak decays. 
Both feeddown and secondary particles
tend to be produced with lower $p_T$ than the corresponding primary particles.
Simulations show that the difference in contamination of $\pi^{+}$ and $\pi^{-}$
by positrons and electrons,
which cannot be rejected within the resolution of the $dE/dx$ measurement,
is negligible.

The acceptance for secondary and feeddown particles is limited
to those produced within 10~cm radial distance from the primary collision vertex,
as accepted tracks were required to have at least one hit in 
the first two layers of the spectrometer.
The background contamination is further reduced by requiring the particle 
tracks to have $dca_{vtx} < 3.5$~mm. 

\begin{figure}[t]
\centerline{
\epsfig{file=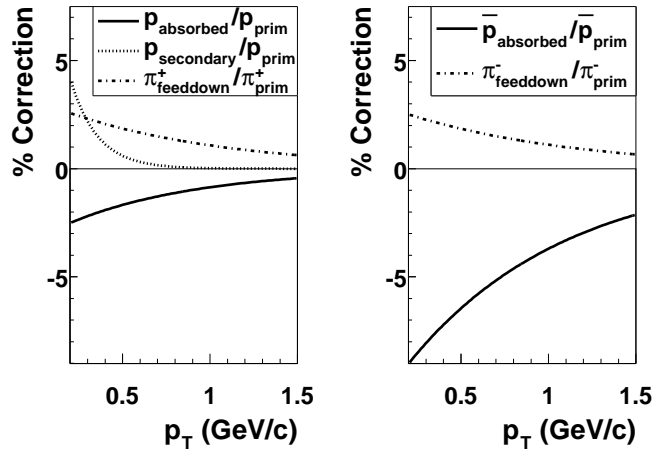,width=9cm}}
\caption{Correction factors for particle multiplicities as a function of transverse 
momentum $p_T$ for positive and negative particles. Correction factors of less
than 1\% and the $p_T$-independent feeddown correction for $\langle \overline{p} \rangle/\langle p \rangle$
are not shown.}
\label{corrpt}
\end{figure}

As HIJING reproduces the total charged particle multiplicity near mid-rapidity
to within 10\%, the simulation should give 
an accurate estimate of the background  from secondary interactions.
The correction factors are shown in Fig.~\ref{corrpt} as a function
of transverse momentum. The net correction is 0.01 for 
$\langle \overline{p} \rangle / \langle p \rangle$.
For pions and kaons the contribution from secondary particles
is much less than 1\%.
The difference in absorption for protons and antiprotons in the
detector material was studied based on GEANT simulations. The fraction
of absorbed protons and antiprotons as a function of $p_T$ is also shown in
Fig.~\ref{corrpt}. From this correction, which averages to 1.7\% for the
protons and 6.3\% for the antiprotons, a correction to 
$\langle \overline{p} \rangle / \langle p \rangle$ of 0.04 is obtained.
The contribution from feeddown particles can not be modelled as precisely, 
as the absolute yield of strange hadrons has not yet 
been measured at RHIC.  
\begin{table}[b]
\caption{Number of accepted events, 
uncorrected number of identified particles and uncorrected $\langle p_T \rangle$ 
for each magnetic field polarity.
\label{table1}}
\begin{tabular}{c|l|c|l|c}
             & BPLUS        & & BMINUS     & \\
             & 26509 events & & 41850 events & \\
\tableline
& \# of particles & $\langle p_T \rangle$ (MeV/c) & \# of particles & $\langle p_T \rangle$ (MeV/c) \\
\tableline
$\pi^+$        & 6208  & $412 \pm 1$  & 23223 & $254 \pm 1$ \\
$\pi^-$        & 14783 & $253 \pm 1$  & 9679  & $410 \pm 1$  \\
$K^+$          & 136   & $358 \pm 4$  & 256   & $248 \pm 4$ \\
$K^-$          & 146   & $255 \pm 5$  & 198   & $366 \pm 4$ \\
p              & 223   & $581 \pm 10$ & 331   & $475 \pm 8$ \\
$\overline{p}$ & 109   & $481 \pm 12$ & 218   & $560 \pm 8$ \\
\end{tabular}
\end{table}

Simulations show that the feeddown correction for $\langle K^- \rangle / \langle K^+ \rangle$ and
$\langle \pi^- \rangle / \langle \pi^- \rangle$  (see Fig.~\ref{corrpt}) is less than
1\%, whereas
feeddown to protons and antiprotons from $\Lambda$ and $\overline{\Lambda}$ is not negligible. 
The feeddown correction to the $\langle \overline{p} \rangle / \langle p \rangle$ ratio 
can be evaluated as a function of the $\langle \overline{\Lambda} \rangle / \langle \Lambda \rangle$ and 
$\langle \Lambda \rangle / \langle p \rangle$ ratios.
Transport model studies, as well as quark counting arguments and 
data from lower energies show that to a  good approximation the following 
relationship holds \cite{zimanyi}:
\begin{eqnarray*}
\frac{\langle \overline{\Lambda} \rangle}{\langle \Lambda \rangle} = 
\frac{\langle K^+ \rangle} {\langle K^- \rangle} \cdot 
\frac{\langle \overline{p} \rangle} {\langle p \rangle}
\end{eqnarray*}
Our data show $\langle K^+ \rangle / \langle K^- \rangle = 1.1$, suggesting 
only a 10\% difference in $\langle \overline{\Lambda} \rangle / \langle \Lambda \rangle$
compared to $\langle \overline{p} \rangle / \langle p \rangle$. 
Using the $dca_{vtx} < 3.5$~mm cut, less than half
of the protons from weak decays are accepted in the analysis.
This further reduces the feeddown correction  to 
$\langle \overline{p} \rangle / \langle p \rangle$.
Using MC simulations, the range of the correction was estimated  by varying the
$\langle \Lambda \rangle / \langle p \rangle$ ratio from 0.2, the value predicted by 
HIJING, to 0.4 and varying
$\langle K^+ \rangle / \langle K^- \rangle$ from 1 to 1.2. The resulting
correction to $\langle \overline{p} \rangle / \langle p \rangle$ ranges from
0 to -0.03, with a most probable value of -0.01.

After corrections we find the following ratios within our acceptance:
\begin{eqnarray*}
&\langle \pi^- \rangle / \langle \pi^+ \rangle & =  1.00 \pm 0.01\mbox{(stat.)}\pm 0.02\mbox{(syst.)}\\
&\langle K^- \rangle/ \langle K^+ \rangle & =  0.91\pm 0.07\mbox{(stat.)}\pm 0.06\mbox{(syst.)}\\
&\langle \overline{p} \rangle / \langle p \rangle & =  0.60\pm 0.04\mbox{(stat.)}\pm 0.06\mbox{(syst.)}
\end{eqnarray*}
For all three particle species the average transverse momentum of 
antiparticles and particles agrees within statistical error (see Table~\ref{table1}), indicating 
no strong dependence of the ratios on transverse momentum. The results are in good 
agreement with data recently presented by the BRAHMS, PHENIX and STAR collaborations \cite{qm01}. 
%This is particularly remarkable for 
%antiprotons and protons, as one third of the protons carry baryon quantum number that is 
%transported from beam-rapidity towards mid-rapidity in the collision process.
\begin{figure}[t]
\centerline{
\epsfig{file=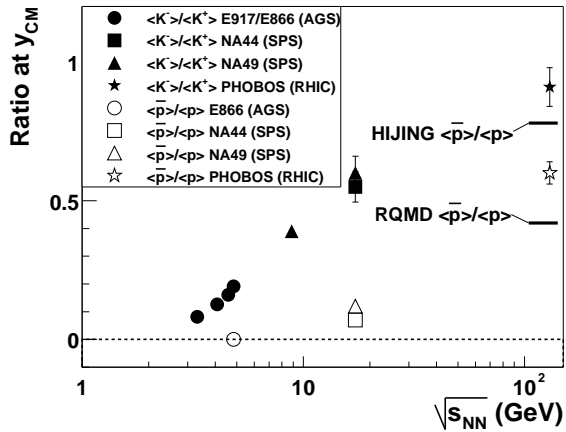,width=8cm}}
\caption{$\langle K^- \rangle/ \langle K^+ \rangle$ and $\langle \overline{p} \rangle / \langle p \rangle$ ratios as a function of $\sqrt{s}$ for 
nucleus-nucleus collisions, in comparison with
predictions from the HIJING and RQMD models. Only statistical errors are shown.}
\label{ratio_roots}
\end{figure}
In Fig.~\ref{ratio_roots} our results are compared to lower energy data 
\cite{e917,na49,na44} and calculations using the HIJING and RQMD\cite{rqmd} 
microscopic transport models. 
The values for the $\langle K^- \rangle/ \langle K^+ \rangle$ 
and $\langle \overline{p} \rangle / \langle p \rangle$ ratios 
are significantly higher than at lower energies.
Using the same parameters that successfully described the charged particle 
multiplicity density in Au+Au collisions at RHIC \cite{phobosprl},
HIJING overestimates $\langle \overline{p} \rangle / \langle p \rangle $ by 0.18. 
Clearly, the particle ratio data provide additional constraints for microscopic models.  
A better description may require a modification in the baryon
number stopping 
mechanism, baryon-pair production or antibaryon annihilation. 
An indication of possible 
mechanisms is given by the comparison with RQMD, which includes  
rescattering of produced hadrons and predicts a value for 
$\langle \overline{p} \rangle / \langle p \rangle$ that is 0.18 below our data.
\begin{figure}[t]
\centerline{
\epsfig{file=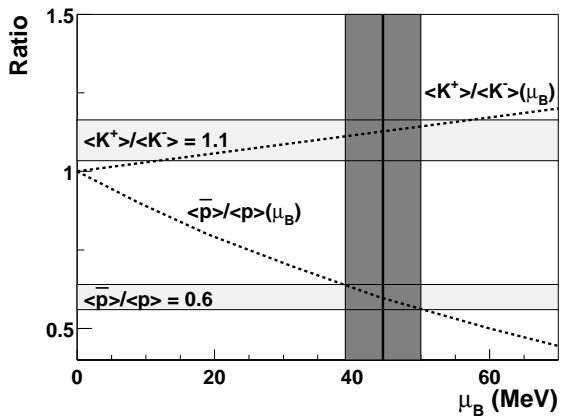,width=8cm}}
\caption{Statistical model calculation  (dotted lines) of $\langle K^- \rangle/ \langle K^+ \rangle$ and
$\langle \overline{p} \rangle / \langle p \rangle$ as a function of $\mu_B$ by Redlich et al. 
The horizontal bands show the ratios observed in the data (statistical errors only).
The vertical shaded area indicates the allowed region in $\mu_B$. }
\label{redlich}
\end{figure}

Finally, we estimate the baryo-chemical potential $\mu_B$ 
using a statistical model calculation \cite{redlich_qm2001} shown
in Fig.~\ref{redlich}.  
For a realistic range of freeze-out temperatures of 160 to 170 MeV, both 
$\langle K^+ \rangle/ \langle K^- \rangle$ and 
$\langle \overline{p} \rangle / \langle p \rangle$
are consistent with $\mu_B = 45 \pm 5$~MeV. 
This is much
lower than the value of $\mu_B = 240 - 270$~MeV \cite{becattini,pbm} obtained in statistical model 
fits to Pb+Pb data at $\sqrt{s_{_{NN}}}= 17.2$~GeV, showing a closer but 
not yet complete approach to a baryon-free regime at RHIC. 

%We acknowledge the generous support of the entire RHIC project personnel, C-A
%and Chemistry Departments at BNL. We thank Fermilab and CERN for help in
%silicon detector assembly. We thank the MIT School of Science and LNS for
%financial support. 
This work was partially supported by US DoE grants DE-AC02-98CH10886,
DE-FG02-93ER-404802, DE-FC02-94ER40818, DE-FG02-94ER40865, DE-FG02-99ER41099, W-31-109-ENG-38.
NSF grants 9603486, 9722606 and 0072204. The Polish groups were partially supported by KBN grant 2 P03B
04916. The NCU group was partially supported by NSC of Taiwan under 
contract NSC 89-2112-M-008-024.

\end{document}